\documentclass[preprint,showpacs,preprintnumbers,amsmath,amssymb,
superscriptaddress]{revtex4}

\usepackage{graphicx}
\usepackage{dcolumn}
\usepackage{bm}

\begin{document}


\title{Investigation of $\beta\beta$ decay in $^{150}$Nd and 
 $^{148}$Nd to the excited states of daughter nuclei}

\author{A.S.~Barabash} \email{barabash@itep.ru}
\affiliation{Institute of Theoretical and Experimental Physics, B.\
Cheremushkinskaya 25, 117259 Moscow, Russian Federation}
\author{Ph.~Hubert}%
\affiliation{Centre d'Etudes Nucl\'eaires,
IN2P3-CNRS et Universit\'e de Bordeaux, 33170 Gradignan, France}
\author{A.~Nachab}
\affiliation{Centre d'Etudes Nucl\'eaires,
IN2P3-CNRS et Universit\'e de Bordeaux, 33170 Gradignan, France}
\author{V.I.~Umatov}
\affiliation{Institute of Theoretical and Experimental Physics, B.\
Cheremushkinskaya 25, 117259 Moscow, Russian Federation}

\date{\today}

\begin{abstract}
Double beta decay of $^{150}$Nd and $^{148}$Nd to the excited states 
of daughter nuclei
have been studied using a 400 cm$^3$ low-background HPGe detector and 
an external 
source consisting of 3046 g of natural Nd$_2$O$_3$ powder. The half-life for 
the two-neutrino 
double beta decay of $^{150}$Nd to the excited 0$^+_1$ state in $^{150}$Sm is 
measured to be 
$T_{1/2}=[1.33^{+0.36}_{-0.23}(stat)^{+0.27}_{-0.13}(syst)]\cdot 10^{20}$ y. 
For other $(0\nu + 2\nu)$ transitions to the
2$^+_1$, 2$^+_2$, 2$^+_3$, and 0$^+_2$ levels in $^{150}$Sm,  limits 
are obtained at the level of 
$\sim (2-8)\cdot 10^{20}$ y. In the case of $^{148}$Nd only limits for the 
$(0\nu + 2\nu)$ transitions 
to the 2$^+_1$, 0$^+_1$, and 2$^+_2$ excited states in $^{148}$Sm were obtained 
and are at the level of 
$\sim (4-8)\cdot 10^{20}$ y.  
\end{abstract}

\maketitle

\section{Introduction}
The experiments with solar, atmospheric, reactor and accelerator neutrinos 
have provided compelling 
evidence for the existence of neutrino oscillations driven by nonzero 
neutrino masses and neutrino mixing (see recent reviews 
\cite{VAL08,BIL07,MOH06} and references therein).
These results are impressive proof that neutrinos have a nonzero mass. 
However, the experiments studying neutrino oscillations are not sensitive 
to the nature of the neutrino mass (Dirac or Majorana) 
and provide no information on the absolute scale of the neutrino masses, 
since such experiments are sensitive only to the difference of the masses, 
$\Delta m^2$. The detection and study of neutrinoless double beta 
($0\nu\beta\beta$)
decay may clarify the following problems of neutrino physics (see discussions 
in \cite{MOH05,PAS06,PAS08}: (i) neutrino nature: 
whether the neutrino is a Dirac or a Majorana particle, (ii) absolute neutrino 
mass scale (a measurement or a limit on $m_1$), (iii) the type of neutrino mass
 hierarchy (normal, inverted, or quasidegenerate), (iv) CP violation in the 
lepton sector (measurement of the Majorana CP-violating phases).

Double
beta decay with the emission of two neutrinos $(2\nu\beta\beta)$ is
an allowed process of second order in the Standard Model. 
The $2\nu\beta\beta$ decays provide the
possibility of an experimental determination  of the 
nuclear matrix  elements (NME) involved
in the double beta decay  processes.  This leads to the development of
theoretical  schemes for NME calculations  both in
connection   with  the   $2\nu\beta\beta$  decays   as  well   as  the
$0\nu\beta\beta$ decays (see, for example, \cite{ROD06,KOR07,KOR07a}).  
At present,
$2\nu\beta\beta$ decay  to  the ground  state  of the  final
daughter   nucleus  has been   measured  for   ten   nuclei:  $^{48}$Ca,
$^{76}$Ge, $^{82}$Se, $^{96}$Zr, $^{100}$Mo, $^{116}$Cd, $^{128}$Te, 
$^{130}$Te, 
$^{150}$Nd  and $^{238}$U  (a review of the results is given in
Refs.~\cite{BAR08,BAR06a,AVI08}).

The $\beta\beta$  decay can proceed through transitions  to the ground
state as  well as to various  excited states of  the daughter nucleus.
Studies of  the latter transitions  allow one to obtain supplementary
information about  $\beta\beta$ decay.  
Because  of smaller transition
energies,  the  probabilities  for $\beta\beta$-decay  transitions  to
excited  states  are   substantially  suppressed  in  comparison  with
transitions to the ground state.  
But as it was shown in Ref.~\cite{BAR90}, by
using   low-background  facilities   utilizing  High Purity Germanium
(HPGe)   detectors,  the 
$2\nu\beta\beta$ decay  to the $0^+_1$  level in the  daughter nucleus
may  be detected for  such nuclei  as $^{100}$Mo,  $^{96}$Zr, and
$^{150}$Nd.  For these isotopes  the energies involved in the
$\beta\beta$ transitions are large (1903,  2202, and
2627~keV, 
respectively),  and  the  expected  half-lives  are of  the  order  of
$10^{20} - 10^{21}$~y.  The  required sensitivity was reached for
$^{100}$Mo in  four independent
experiments~\cite{BAR95,BAR99,DEB01,HOR06,ARN07}. It was also obtained in
$^{150}$Nd~\cite{BAR04}.
Recently    additional    isotopes ($^{82}$Se,    $^{130}$Te,
$^{116}$Cd, and $^{76}$Ge) have also become of interest to
studies of the $2\nu\beta\beta$ decay to the $0^+_1$ level (see reviews
in Refs.~\cite{BAR00,BAR04a,BAR07}).

The $0\nu\beta\beta$ transition  to excited
states of  daughter nuclei provide a clear-cut signature  of 
such  decays, and is
worthy of a special  note here. In addition to two electrons with fixed
total energy, one  ($0^+ \to 2^+$ transition) or two ($0^+
\to  0^+$ transition) photons appear,  with their energies  being strictly  fixed. 
In  a hypothetical experiment  detecting all decay products  with a
high efficiency and a high  energy resolution, the background can be
reduced to nearly zero.  This zero background idea will be the goal of
future experiments featuring a large mass of the $\beta\beta$ sample
(Refs.~\cite{BAR04a,BAR00,SUH00}).     
In Ref.~\cite{SIM02} it  was mentioned that detection  of this transition
will   give    us  the additional   possibility of distinguishing the
$0\nu\beta\beta$ mechanisms (the light and heavy Majorana neutrino
exchange mechanisms, the trilinear R-parity breaking mechanisms etc.).
So   the   search  for   $\beta\beta$  
transitions to the excited states has its own special interest.

\begin{figure*}
\begin{center}
\includegraphics[width=12.8cm]{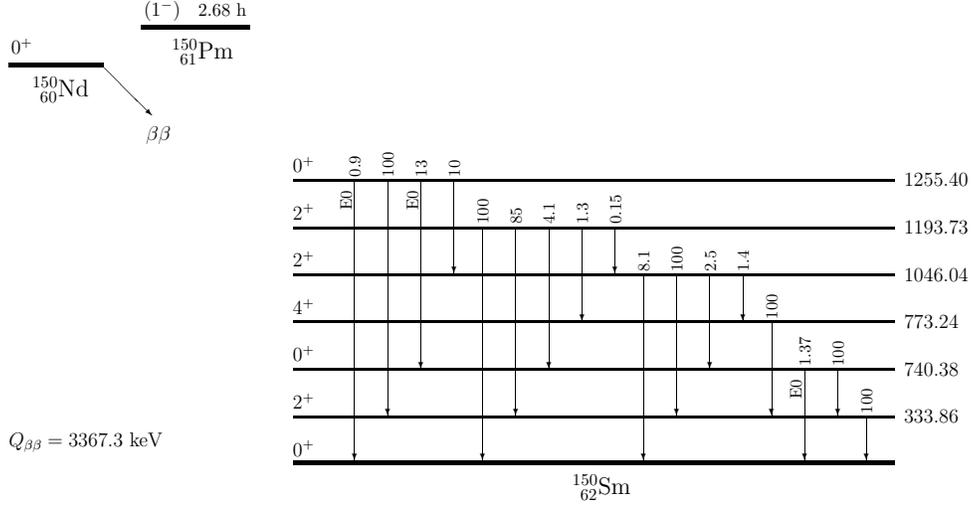}
\caption{\label{fig:fig1}Decay scheme of $^{150}$Nd are taken 
from \cite{DER95}. 
Only the investigated levels of $0^+$ and $2^+$ and levels associated 
with transitions
to the first ones are shown. Relative branching ratios from each level are 
presented.}  
\end{center}
\end{figure*}

\begin{figure}
\begin{center}
\includegraphics[width=7.0cm]{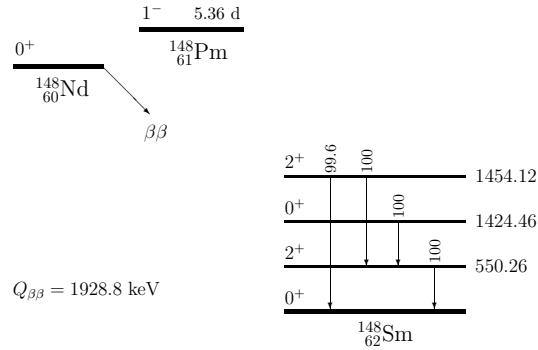}
\caption{\label{fig:fig2}Decay scheme of $^{148}$Nd \cite{BHA00}. 
Only the investigated levels of $0^+$ and $2^+$ are shown. The relative 
branching from each level are also given.}  
\end{center}
\end{figure}

In this article, results of an experimental investigation
of the $\beta\beta$ decay of $^{150}$Nd and $^{148}$Nd to the excited states in
$^{150}$Sm and  $^{148}$Sm are presented. 
The decay schemes for the triplets $^{150}$Nd - $^{150}$Pm - $^{150}$Sm  
\cite{DER95}
and $^{148}$Nd - $^{148}$Pm - $^{148}$Sm \cite{BHA00}
are shown in Fig.~\ref{fig:fig1} and Fig.~\ref{fig:fig2} respectively. 
 The search for $\beta\beta$ transitions of $^{150}$Nd and $^{148}$Nd
to excited states has been carried out using a
HPGe detector to look for $\gamma$-ray lines corresponding to their
decay schemes. A preliminary result for $\beta\beta$ decay of $^{150}$Nd 
to the $0^+_1$ 
excited state of $^{150}$Sm
was published in Ref. \cite{BAR04}. 

\section{Experimental study}

The experimental work was performed in the Modane Underground
Laboratory (depth of 4800 m w.e.). A 400 cm$^3$ low-background HPGe detector 
was used to measure a 3046 g sample of Nd$_2$O$_3$ powder 
in a special Marinelli delrin box which was placed on the detector endcap.
Taking into account the natural abundance there are  
153 g of $^{150}$Nd (5.64\%) and 154 g of $^{148}$Nd (5.76\%) in the sample. 
Data were collected for 11320.5 h.

The Ge spectrometer is composed of a p-type crystal.  
The cryostat, endcap, and the other mechanical parts have been made of 
a very pure 
Al-Si alloy. The cryostat has a J-type geometry to shield the crystal from 
radioactive impurities in the dewar. The passive shielding consisted of 
2 cm of archeological lead, 10 cm of OFHC copper, and 15 cm of ordinary lead. To 
remove $^{222}$Rn gas, one of the main sources of the background, a special 
effort was made to minimize the free space near the detector. In addition, 
the passive shielding was enclosed in an aluminum box flushed with high-purity
 nitrogen.

The electronics consisted of currently available spectrometric amplifiers and
 a 8192 channel ADC. The energy calibration was adjusted to cover the energy 
range from 50 keV to 3.5 MeV, and the energy resolution was
 2.0 keV for the 1332-keV line of $^{60}$Co. The electronics were stable during
 the experiment due to the constant conditions in the laboratory (temperature 
of $\approx 23^\circ$ C, hygrometric degree of $\approx 50$\%).  A daily check
 on the apparatus assured that the counting rate was statistically constant.

The current data of accepted values for different
isotopes published in Nuclear Data Sheets were used
for analysis of the energy spectrum. 
The photon detection efficiency for each investigated process 
has been calculated
with the CERN Monte Carlo code GEANT 3.21.  
Special calibration
measurements with radioactive sources and powders containing 
well-known $^{226}$Ra activities confirmed that the accuracy of these 
efficiencies is about 10\%.

The dominate detector backgrounds come from natural $^{40}$K,
radioactive chains of $^{232}$Th and $^{235,238}$U, man-made and/or
cosmogenic activities of $^{137}$Cs and $^{60}$Co.
 The sample was found to have a large activity
of $^{40}$K (46.3 mBq/kg).
Additionally long-lived radioactive impurities were observed in the sample, 
but with much weaker activities, i.e.  $^{137}$Cs (0.089 mBq/kg),  $^{176}$Lu
(0.450 mBq/kg),  $^{138}$La (0.068 mBq/kg),  $^{133}$Ba (0.155 mBq/kg), etc.
 In our case the most important isotopes 
contributing to energy ranges of the investigated transitions are 
$^{214}$Bi (1.15 mBq/kg), $^{228}$Ac (0.93 mBq/kg), 
$^{227}$Ac (0.62 mBq/kg), and their daughters.

\begin{figure}
\begin{center}
\includegraphics[width=8.6cm]{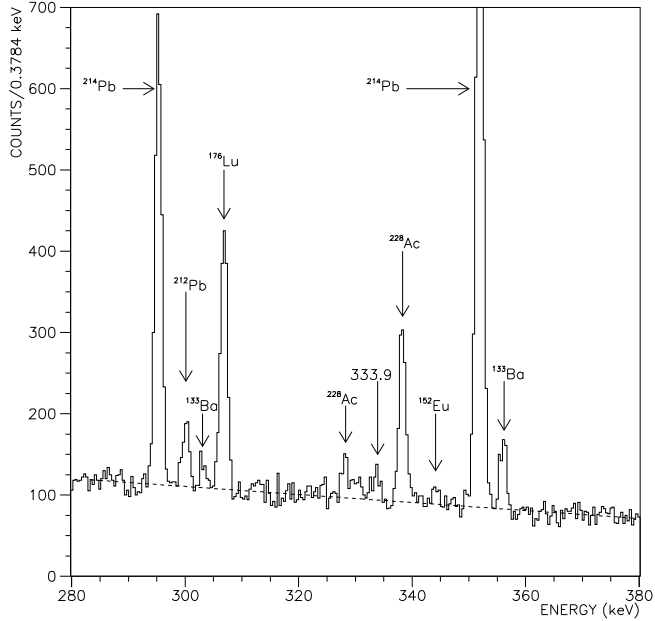}
\caption{\label{fig:fig3}Energy spectrum with natural Nd$_2$O$_3$ powder of 
observed $\gamma$-rays in the range [280-380] keV.}  
\end{center}
\end{figure}

\begin{figure}
\begin{center}
\includegraphics[width=8.6cm]{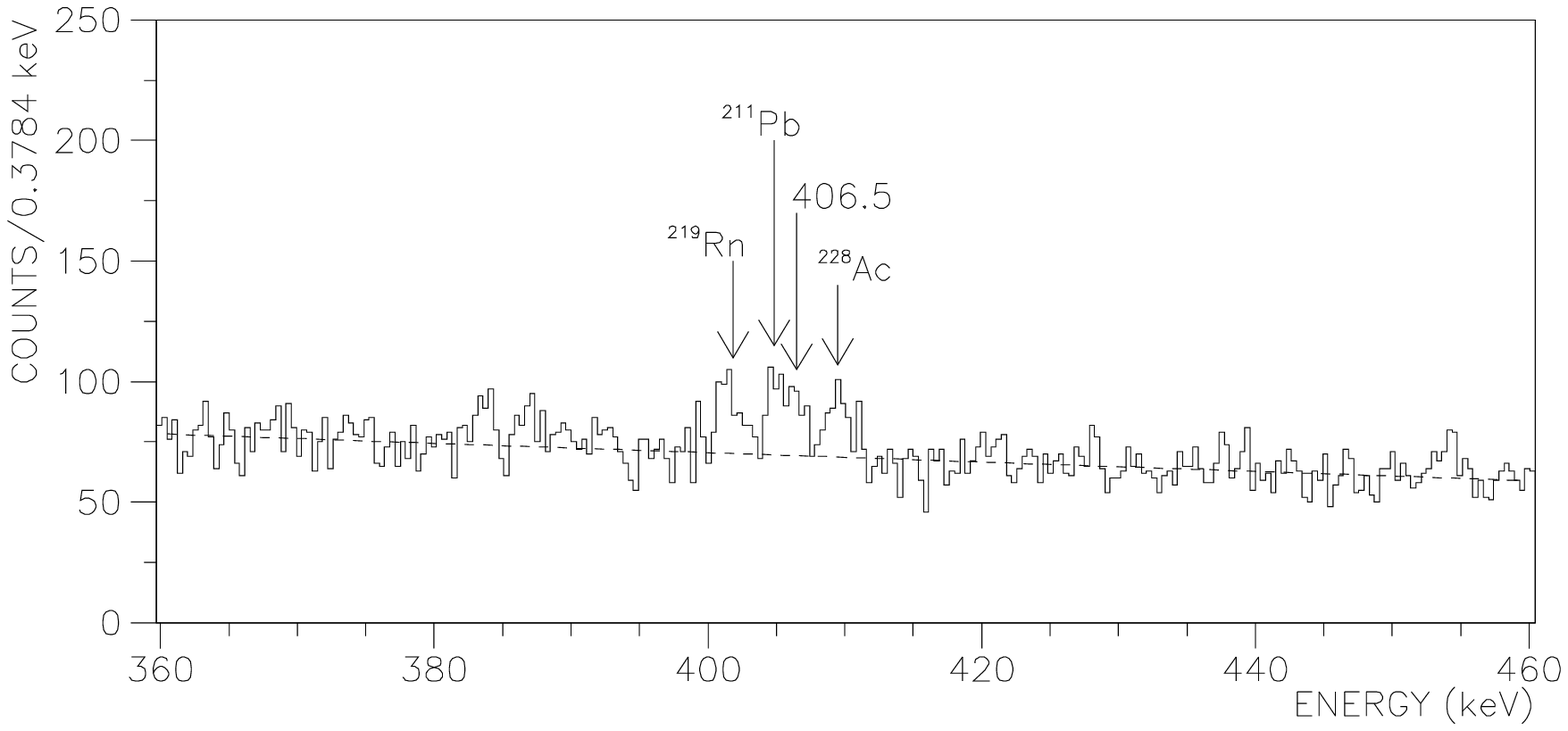}
\caption{\label{fig:fig4}Energy spectrum with natural Nd$_2$O$_3$ powder of 
observed $\gamma$-rays in the range [360-460] keV.}  
\label{fig_2}
\end{center}
\end{figure}

\begin{figure}
\begin{center}
\includegraphics[width=8.6cm]{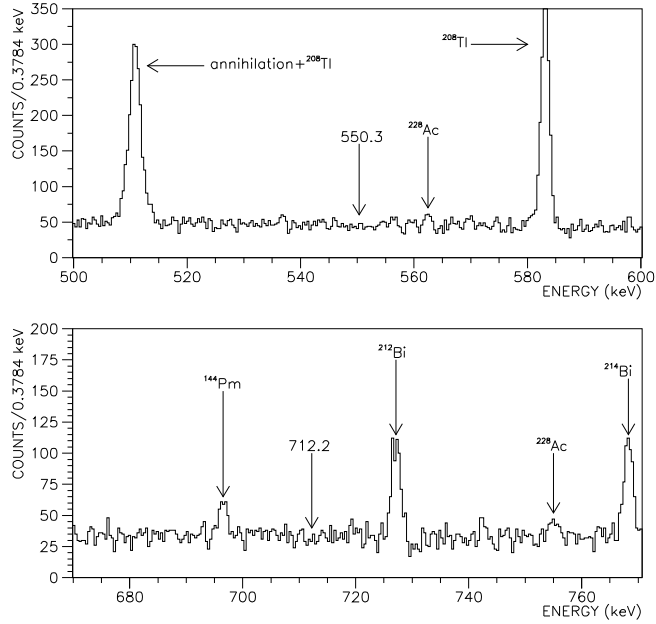}
\caption{\label{fig:fig5}Energy spectrum with natural Nd$_2$O$_3$ 
powder in the ranges of investigated 
 $\gamma$-rays ([500-600]  and [670-770] keV).}  
\end{center}
\end{figure}

\begin{figure}[h]
\begin{center}
\includegraphics[width=8.6cm]{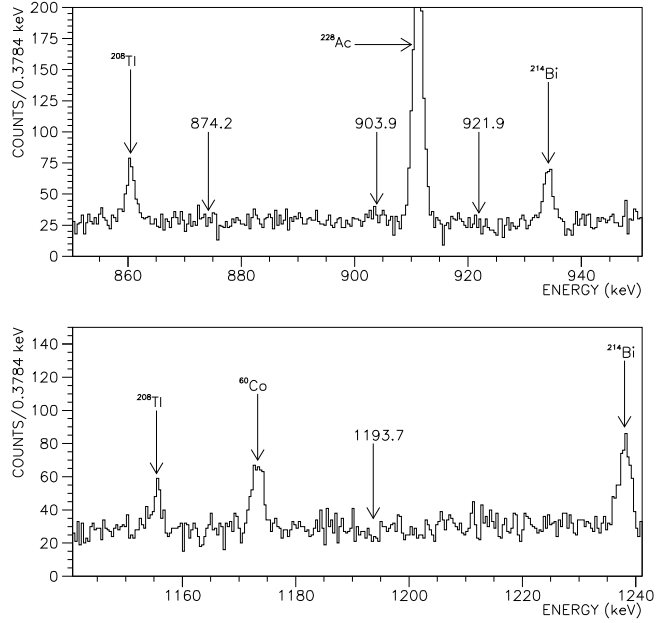}
\caption{\label{fig:fig6}Energy spectrum with natural Nd$_2$O$_3$ 
powder for $\gamma$-rays in the ranges [850-950]  and [1140-1240] keV.}  
\end{center}
\end{figure}

Figures \ref{fig:fig3}, \ref{fig:fig4}, \ref{fig:fig5}, and \ref{fig:fig6}
 show the energy spectra in the ranges of interest.

\section{Analysis and results}

\subsection{Search for $\beta\beta$ processes in $^{150}$Nd}

Double beta decays of $^{150}$Nd to 2$^+_1$ (333.86 keV), 0$^+_1$ (740.38 keV), 
2$^+_2$ (1046.04 keV), 2$^+_3$ (1193.73 keV) and 
0$^+_2$ (1255.40 keV) levels in $^{150}$Sm
have been investigated. 

\subsubsection{Decay to the 0$^+_1$ excited state}

The transition is accompanied by two $\gamma$-rays with energies of 
333.9 keV and
406.5 keV. The detection photopeak efficiencies are equal to 2.30\% 
at 333.9 keV and
2.29\% at 406.5 keV.
Fig.~\ref{fig:fig3} and Fig.~\ref{fig:fig4} show the energy spectrum 
in the ranges of interest.
As one can see there is an excess of events above the averaged continuous 
background
at the investigated energies. Isotopes of natural radioactivity ($^{211}$Pb,
$^{214}$Bi, $^{227}$Th, and $^{228}$Ac), found in the spectrum, have 
$\gamma$-lines near these energies.
$^{214}$Bi contributes 
to both investigated ranges through $\gamma$-rays with energies of
 333.37 keV (0.065\%) and 334.78 keV (0.036\%) for the
333.9-keV peak, and 405.72 keV (0.169\%) for the 406.5-keV peak. 
$^{228}$Ac touches 
the 333.9-keV peak range with its $\gamma$ (332.37 keV, 0.40\%).
Appropriate $\gamma$-rays from $^{227}$Th and $^{211}$Pb are 334.37 keV (1.54\%) 
and 404.853 keV (3.78\%), respectively. Both of the mentioned isotopes are 
daughters of $^{227}$Ac ($T_{1/2}=21.772$ y; $^{235}$U chain). Activity of
$^{227}$Ac was estimated using the most intensive $\gamma$-lines of its daughters,
$^{227}$Th (235.96 keV, 12.9\%; 256.23 keV, 7.0\%), $^{223}$Ra (269.46 keV, 13.9\%)
and $^{219}$Rn (271.23 keV, 10.8\%; 401.81 keV, 6.6\%). 

There is also the cosmogenic isotope, $^{150}$Eu ($T_{1/2}=36.9$ y), 
which decays to
$^{150}$Sm, with $\gamma$-rays of 333.9 keV (96\%), 406.5 keV (0.14\%),
439.4 keV (80\%), 584.3 keV (52.6\%). The line at 439.4
keV is within one standard deviation of the continuous background, 
therefore it can
be taken into account as a systematic error.

Table~\ref{tab:table1} presents the results of the analysis for the 
two peak energy ranges 
being studied. A peak's shape is described by a gaussian with a standard 
deviation of 0.58 keV at 333.9 keV and 0.61 keV at 406.5 keV. 
For the analysis a peak's range is
taken to within approximately four standard deviations (E$\pm 2\sigma$), i.e., 
94.82\% of full peak at 333.9 keV and 92.75\% of full peak at 406.5 keV.  
As one can see there is an excess
of events for each peak. Summing the two peaks we obtain a signal
of ($177.5\pm 37.6$) events, corresponding to a half-life of $^{150}$Nd 
to the first $0^+$ 
excited state of $^{150}$Sm given by 
$T_{1/2}=[1.33^{+0.36}_{-0.23}(stat)^{+0.23}_{-0.13}(syst)]\cdot 10^{20}$ y. 

\begin{table*}
\caption{\label{tab:table1}Analysis of events in the energy ranges of the 
peaks under study.}
\begin{ruledtabular}
\begin{tabular}{llcrc}
\rule[-1mm]{0mm}{5mm}
Peak (keV) & & $333.9$ & & $406.5$ \\ \hline 
\rule[-1mm]{0mm}{5mm}
Number of events & & 779 & & 603 \\ 
Continuous background & & {$656.4\pm 1.6 $} & & {$484.6\pm 1.2$} \\ 
\rule[-1mm]{0mm}{5mm}
Isotope & $^{214}$Bi(333.37;334.78) & $7.9\pm 2.2$ & $^{214}$Bi(405.72) 
& $9.3\pm 2.1$ \\
contributions & $^{227}$Th(334.37) & $30.7\pm 4.7$ & $^{211}$Pb(404.85) &
  $10.2\pm 1.1$ \\
 & $^{228}$Ac(332.37) & $5.6\pm 0.7$ \\ 
\rule[-1mm]{0mm}{5mm}
Excess of events & & $78.5\pm 28.4$ & & $99.0\pm 24.7$ \\ 
\end{tabular}
\end{ruledtabular}

\end{table*}

The primary systematics come from the GEANT calculations (10\%), 
continuous background estimation (2.6\%), and the possible contribution 
of $^{150}$Eu in the 333.9-keV peak (8.5\%).

Previous experiments gave only limits on this transition, 
$>1\cdot 10^{20}$ y \cite{ARP99} and $>1.5\cdot 10^{20}$ y \cite{KLI02}.
Taking into account all errors, our "positive" result is not in contradiction 
with the previous limits.

\subsubsection{\label{sec:level3}Decay to the 2$^+_1$ excited state}

To search for this transition one has to look for a $\gamma$-ray with an energy 
of 333.9 keV.  The detection efficiency is 2.60\%. The analysis given above 
shows
 that the excess of events at 333.9 keV is mainly due to the double beta decay 
of $^{150}$Nd to the 0$^+_1$ excited state of $^{150}$Sm. So one can only give 
the lower half-life limit on the transition to the 2$^+_1$ excited state
of $^{150}$Sm. The limit has been calculated using
the likelihood function described in Refs. 
\onlinecite{BAR96,BAR96a} which takes into account all the peaks 
identified above as background.
This result together with available data on $\beta\beta$ decay of $^{150}$Nd 
from other experimental works are presented in Table II.

\subsubsection{Decays to the 2$^+_2$, 2$^+_3$ and 
0$^+_2$ excited states}

To search for these transitions one has to look for $\gamma$-rays with 
energies of 
712.2, 921.5 and 1193.7 keV (Fig.~\ref{fig:fig1}).  
As one can see from figures 
\ref{fig:fig5} and \ref{fig:fig6},
 there are no statistically significant peaks at
these energies. Using the same technique as above \cite{BAR96,BAR96a} the lower 
half-life limits are found within $(4.7-8.0) \cdot 10^{20}$ y for the 
transitions (Table~\ref{tab:table2}). 
Table~\ref{tab:table2} also presents other valuable data on 
these transitions.

\begin{table*}
\caption{\label{tab:table2}Experimental results for $(0\nu+2\nu)\beta\beta$ 
decay of $^{150}$Nd 
to the excited states of $^{150}$Sm. All limits are given at the 90\% C.L.}

\begin{ruledtabular}
\begin{tabular}{cccl}
Excited state & Energy of $\gamma$-rays & 
\multicolumn{2}{c}{$(T^{0\nu+2\nu}_{1/2})_{exp}$ ($10^{20}$ y)} \\ 
 \cline{3-4}
 & (efficiency) & this work & other works \\ 
\hline
$2^+_1 (333.86)$   & 333.9  (2.60\%) & $ > 2.2 $ & $ > 0.91 $\cite{ARP99}  \\
                   &                 &           & 
$ > 24 $\footnotemark[1] \cite{NEMO08}  \\
$0^+_1 (740.38)$  & 333.9  (2.30\%) & 
$ =[1.33^{+0.36}_{-0.23}(stat)^{+0.27}_{-0.13}(syst)]$\footnotemark[2] & 
$ > 1.5 $ \cite{KLI02}  \\
                   & 406.5  (2.29\%) &  & $ > 1 $ \cite{ARP99} \\ 
                   &                 &  & 
$ > 2.4$\footnotemark[1]\cite{NEMO08}  \\                  
$2^+_2 (1046.04)$  & 712.2  (1.78\%) & $ > 8.0$ & $ > 1.4 $ \cite{ARP96}  \\
$2^+_3 (1193.73)$  & 1193.7 (0.95\%) & $ > 5.4$ & $ > 0.027 $ \cite{BEL82} \\
$0^+_2 (1255.40)$  & 921.5  (1.45\%) & $ > 4.7$ & $ > 2 $ \cite{ARP99}  \\
\end{tabular}
\end{ruledtabular}
\footnotetext[1]{Only 0$\nu$ decay mode}
\footnotetext[2]{Half-life value for $ 2\nu$ decay (see text for the details)}
\end{table*}

\subsection{Search for $\beta\beta$ processes in $^{148}$Nd}

A search for the double beta decays of $^{148}$Nd to the 2$^+_1$, 0$^+_1$, and 
2$^+_2$ excited states of $^{148}$Sm was carried out by looking for 
$\gamma$-rays with energies of 550.3, 874.2, and 903.9 keV accompanying 
these transitions (Fig.~\ref{fig:fig2}).
Figures \ref{fig:fig5} and \ref{fig:fig6} show no statistically 
significant peaks at these energies. 
The lower half-life limits reported in Table~\ref{tab:table3} have 
been calculated using
the same procedure as in section \ref{sec:level3}.   
Available data on $\beta\beta$ decay of $^{148}$Nd from other 
experimental works are also presented in Table~\ref{tab:table3}.

\begin{table}
\caption{\label{tab:table3}Experimental results for 
$(0\nu+2\nu)\beta\beta$ decay of $^{148}$Nd 
to the excited states of $^{148}$Sm. All limits are given at the 90\% C.L.
(See text for details.)}
\begin{ruledtabular}
\begin{tabular}{cccc}
Excited state & Energy of $\gamma$-rays & 
\multicolumn{2}{c}{$(T^{0\nu+2\nu}_{1/2})_{exp}$ ($10^{20}$ y)} \\ 
 \cline{3-4}
 & (efficiency) & this work & other works \\ \hline
$2^+_1 (550.26)$   & 550.3  (2.36\%) & $ > 6.6 $ & $ > 0.03 $ \cite{BEL82}  \\
$0^+_1 (1424.46)$  & 550.3  (2.16\%) & $ > 7.9 $ & -  \\
                   & 874.2  (1.83\%) &                     \\
$2^+_2 (1454.12)$  & 550.3  (1.11\%) & $ > 3.8$ & $ > 0.027 $\cite{BEL82} \\
                   & 903.9  (0.87\%) &     \\
\end{tabular}
\end{ruledtabular}
\end{table}

\section{Discussion} 

Because the technique used in the present work does not allow for a distinction 
between $0\nu\beta\beta$ and $2\nu\beta\beta$ decay, our 
result for double beta decay of $^{150}$Nd to the excited 0$^+_1$ state in 
$^{150}$Sm is the sum of the $0\nu\beta\beta$ and $2\nu\beta\beta$ processes. 
However we believe that we detected only the $2\nu\beta\beta$ decay. 
This conclusion is supported by two arguments. First, in the recent NEMO paper 
\cite{NEMO08} the limit on $0\nu\beta\beta$ decay of $^{150}$Nd to the excited 
0$^+_1$ state was established as $2.4\cdot 10^{20}$ y, which is stronger 
than the half-life value obtained here. Second, the experimental limit for the 
$0\nu\beta\beta$ decay of $^{150}$Nd to the ground 0$^+$ state of $^{150}$Sm is
 about two orders of magnitude larger \cite{NEMO08} than the value reported 
here. 
Therefore, considering the reduced phase space factors available for 
the transition to the excited 0$^+_1$ state, it is safe to assume that our 
result for $T_{1/2}$ refers solely to the $2\nu\beta\beta$ decay. As to the 
possible contribution of $(0\nu + 2\nu)\beta\beta$ decay to the $2^+_1$ excited state to the 
peak at 333.9 keV one can conclude that this contribution is small. 
First, the experimental limit on neutrinoless mode 
($T_{1/2} > 2.4\cdot 10^{21}$ y \cite{NEMO08}) 
is much stronger than the obtained "positive" result for the 
decay of $^{150}$Nd to the 0$^+_1$ excited state of $^{150}$Sm. 
And second, the theoretical prediction 
for two neutrino mode 
($T_{1/2} \approx 7.2\cdot 10^{24} - 1.2\cdot 10^{25}$ y \cite{HIR95}) 
is very far from the observed value.   

Using the phase space factor value G = $1.2\cdot 10^{-17} y^{-1}$ 
(for g$_A$ = 1.254) \cite{SUH98}  and the measured half-life 
$T_{1/2}=[1.33^{+0.36}_{-0.23}(stat)^{+0.23}_{-0.13}(syst)]\cdot 10^{20}$ y, 
one 
obtains a NME value for the $2\nu\beta\beta$ transition to the 
0$^+_1$ excited state of ${\it M}_{2\nu}$(0$^+_1$) = $0.025\pm 0.003$ (scaled by 
the electron rest mass).
One can compare this value with the NME value for the $2\nu\beta\beta$ 
transition 
to the ground 
state of $^{150}$Sm, ${\it M}_{2\nu}$(0$^+_{g.s.}$) = 0.033$^{+0.001}_{-0.002}$ (here 
we used the 
average half-life value $T_{1/2}=(7.8\pm 0.7)\cdot 10^{18}$ y from 
\cite{BAR06a}  and 
G = $1.2\cdot 10^{-16} y^{-1}$ (for g$_A$ = 1.254) from \cite{SUH98}). One can 
see that 
${\it M}_{2\nu}$(0$^+_{g.s.})$ is $\sim$ 25\% greater than ${\it M}_{2\nu}$(0$^+_1$). 
Nevertheless, these 
values are very close and taking into account errors and possible uncertainties 
their equality can not be excluded. From the viewpoint of the theory it is 
important to confirm this difference or to rule it out. Future more precise 
measurements for both transitions 
and especially for the transition to the excited state are needed.

For double beta decay of $^{150}$Nd to the 2$^+_1$, 2$^+_2$, and 0$^+_2$ excited 
states of $^{150}$Sm the limits obtained are $\sim$ 2-5 times better than 
the best previous results \cite{ARP99,ARP96,BEL82}. The limit obtained for 
the transition to the 
2$^+_3$ excited state is 200 times better 
than the previous limit \cite{BEL82}. 

The $^{148}$Nd limit for the transition to the 0$^+_1$ excited state 
was obtained for the first time and for transitions to
 2$^+_1$ and 2$^+_2$ states
a sensitivity was achieved that is $\sim$ 200 times better than 
in Ref. \cite{BEL82}.

\section{Conclusion}

Double beta decay of $^{150}$Nd and $^{148}$Nd to the excited states of 
daughter 
nuclei was investigated with a high level of sensitivity. The half-life for 
the 
$2\nu\beta\beta$ decay of $^{150}$Nd to the excited 0$^+_1$ state 
in $^{150}$Sm is measured to be 
$T_{1/2}=[1.33^{+0.36}_{-0.23}(stat)^{+0.27}_{-0.13}(syst)]\cdot 10^{20}$ y. 
The strongest limits
for other transitions were established. The sensitivity of this experiment 
could still be increased by a few times using a pure Nd$_2$O$_3$ (or Nd) 
sample. Also further increases in the sensitivity could be reached using 
an enriched Nd sample and a multicrystal HPGe installation to study 
larger masses of Nd samples.

\section*{Acknowledgements}
The authors would like to thank the Modane Underground Laboratory staff for 
their technical 
assistance in running the experiment. We are very thankful to Prof. S. Sutton 
for his useful remarks. Portions of this work were supported by 
grants from RFBR 
(no 06-02-72553) and CRDF (RUP1-2892-MO-07). 
This work was also supported by the Russian Federal Agency for Atomic Energy.

 --------------------------------------------------------------

\end{document}